\documentclass[twocolumn,aps,prb,showpacs]{revtex4}
\usepackage{bm}
\usepackage{amsmath}
\usepackage{amssymb}
\usepackage{epsfig}
\usepackage{epsf}
\usepackage{array}
\usepackage{color}
\usepackage{graphicx}
\usepackage{epstopdf}
\usepackage{subfigure}
\usepackage{bbm}
\usepackage{hyperref}

\begin{document}
\title{Theory of charged impurities in correlated electron materials: Application to muon spectroscopy of high-Tc superconductors}
\author{Hung T. Dang, Emanuel Gull and Andrew J. Millis}
\affiliation{Department of Physics, Columbia University, 538 West 120th Street, New York, New York 10027, USA}

\begin{abstract}
Single-site and cluster dynamical mean-field methods are used to estimate the response of a doped Mott insulator  to a charged impurity.  The effect of correlations on the Thomas-Fermi screening properties is determined. The charge density, the on-site and near-neighbor spin-spin correlations in the vicinity of the impurity are compared to those far from it.   The theory is used to address the question of the effect of the density perturbation induced by the  muon charge on the local response functions of a high-temperature superconductor. For reasonable values of the background dielectric constant and basic correlation strength,  a muon is shown to lead to an observable perturbation of the local spin dynamics, raising questions about the interpretation of muon-spin-rotation experiments in metallic high-temperature superconductors. 
\end{abstract}

\maketitle

\section{Introduction:} Mobile electrons act to screen a charged impurity. Screening may be understood in terms of two equations: the Poisson equation which relates the electric potential $V(r)$ to the combination of the impurity charge density $\rho_\text{imp}(r)$ and the change $\delta n(r)$ in free charge density, and a constitutive equation which relates $\delta n$ to $V$.  For weakly correlated metals it suffices to linearize the constitutive relation so the screening properties are determined by the  density-density correlation function $\chi$. A locality approximation is  typically appropriate, so that  $\delta n(r)=\chi(r=0,\omega=0)V(r)$ and it  is also reasonable  to write the Poisson equation in its continuum form. These approximations imply that screening in weakly correlated metals is described by  the familiar Thomas-Fermi equations, which lead to an exponential decay of the charge density characterized by the Thomas-Fermi length $\lambda_{TF}=1/\sqrt{4\pi e^2 \chi(r=0,\omega)/\epsilon}$. 

Remarkably, while the ubiquity of impurity effects in correlated electron materials has prompted extensive theoretical studies of the consequences of local disorder for magnetic and superconducting properties,\cite{Alloul09} few results seem to be available for the problem of screening of a charge center in a material with strong electronic correlations. Several effects appear to be important.  First, and most trivially, lattice effects are strong so a discrete version of the Poisson equation must be used. Second, most correlated materials of interest are oxides or organics with high background polarizability. Third, the constitutive relation between potential and density   is likely to be strongly affected by correlation phenomena, which, in particular, will act to reduce the charge response. Fourth, in correlated materials, properties are typically sensitive functions of density, so that  linearization in the magnitude of the density change  may not be appropriate, while charge accumulation or depletion near an impurity may change the physics locally, for example, nucleating  or suppressing local order or fluctuations.

The possibility of local changes in the physics is of particular importance in the context of muon-spin-rotation spectroscopy. In this class of experiments, a positively charged muon with a known initial spin direction is injected into a solid. Coupling to magnetic order or fluctuations causes the spin of the muon to precess before it decays and the amount of precession (and hence some information about the spin fluctuations) can be inferred from the angular distribution of the muon decay products. If the charge of the muon causes a significant perturbation of the electronic properties near the muon site, then the muon does not necessarily measure the intrinsic magnetic dynamics of the material.  

The question of the perturbation imposed by an injected muon has recently arisen in the context of the possible observation of an ``orbital current'' phase in high-temperature superconductors.  Following a prediction of Varma, \cite{Varma97,Varma06} neutron-scattering experiments \cite{Fauque06,Li08,Mook08}  reported evidence of a time-reversal symmetry-breaking phase characterized by local magnetic fields which are non-vanishing but average to zero over a unit cell; however, muon-spin-rotation experiments \cite{Sonier01,Macdougall08,Sonier09} failed to detect the magnetic fields implied by the neutron experiments.   One possible resolution of the discrepancy is that the neutron measurements detect properties of a minority phase. Another possible resolution, proposed by Shekhter \textit{et al.},\cite{Shekhter08} is that  the muon, which carries unit charge, perturbs the local physics strongly enough to destroy the local order detected by neutrons. Reference \onlinecite{Shekhter08} presented a Thomas-Fermi calculation which used a continuum version of the Poisson equation,  a value $\varepsilon\sim 4$  of the dielectric constant rather smaller than the value $\varepsilon\sim 10-15$ generally accepted for oxides and a compressibility which was assumed to be linearizable and   unrenormalized by many-body effects (although  some consequences of the correlations were mentioned). The calculation of Shekhter \textit{et al.}  implies that the muon would constitute a strong perturbation, dramatically changing the doping and the magnetic dynamics. However, the discussion given above implies that the assumptions on which the calculation is based may be questioned.

In this paper we re-examine the issue. We use a tight-binding model description which captures the physics associated with the discreteness of the lattice, we examine the dependence on background dielectric constant, and most importantly we use single-site \cite{Georges96} and cluster \cite{Maier05} dynamical mean-field-based methods  to provide an estimate of the correlation effects on screening and on near-impurity electronic properties. We determine when linearization is appropriate and, where needed, use the full nonlinear (but local) charge response. We compute locally defined quantities which give some insight into changes in spin correlation properties.  While our specific results are obtained for a lattice, doping,  and interaction strength appropriate to high-$T_c$ cuprates, we expect that our methods are more broadly applicable and our qualitative results are relevant to a wider range of systems.  

We find that  if dielectric constants in the physically reasonable range are used, the presence of a unit charge induces density changes which are  a non-negligible fraction of the doping; however, the resulting changes in local magnetic properties are found to be modest, although observable. Further theoretical attention, perhaps using ``LDA+DMFT'' methods \cite{Kotliar06}, should be given to modeling the effects of charge centers in general and muons in particular.

\section{Model}
We approximate the conduction band degrees of freedom as a one-band  Hubbard model
\begin{equation}\label{eqn:Hubbard_model}
H = -\sum_{ij\sigma}t_{ij}c^\dagger_{i\sigma}c_{j\sigma} +\sum_i \left({\bar \mu}+V_i\right)c^\dagger_{i\sigma}c_{i\sigma}
	+ U\sum_i n_{i\uparrow} n_{i\downarrow}
\end{equation}
with  hopping $t_{ij}$, on-site repulsion $U$ and a spatially varying electrochemical potential $\mu_i={\bar \mu}+V_i$ determined  self-consistently (see below)  from the impurity charge and any induced electronic charge.  For the explicit calculations presented in this paper we take a set of planes which are electronically decoupled (no interplane hopping) but coupled via the Coulomb interaction. Each plane is taken to be a two-dimensional square  lattice with nearest-neighbor hopping.  We consider  interactions of the order of the critical value $U_{c2}$ needed to drive a metal insulator transition in a homogeneous bulk system with one electron per site. ${\bar \mu}$, the chemical potential far from the impurity site,  is chosen to produce a carrier density ${\bar n}$ corresponding to a hole doping of  $\delta=1-{\bar n} = 0.1$ corresponding approximately to the doping level at which pseudogap and magnetic effects occur in the high-$T_c$ cuprates. 

\begin{figure}
\includegraphics[width=0.85\columnwidth]{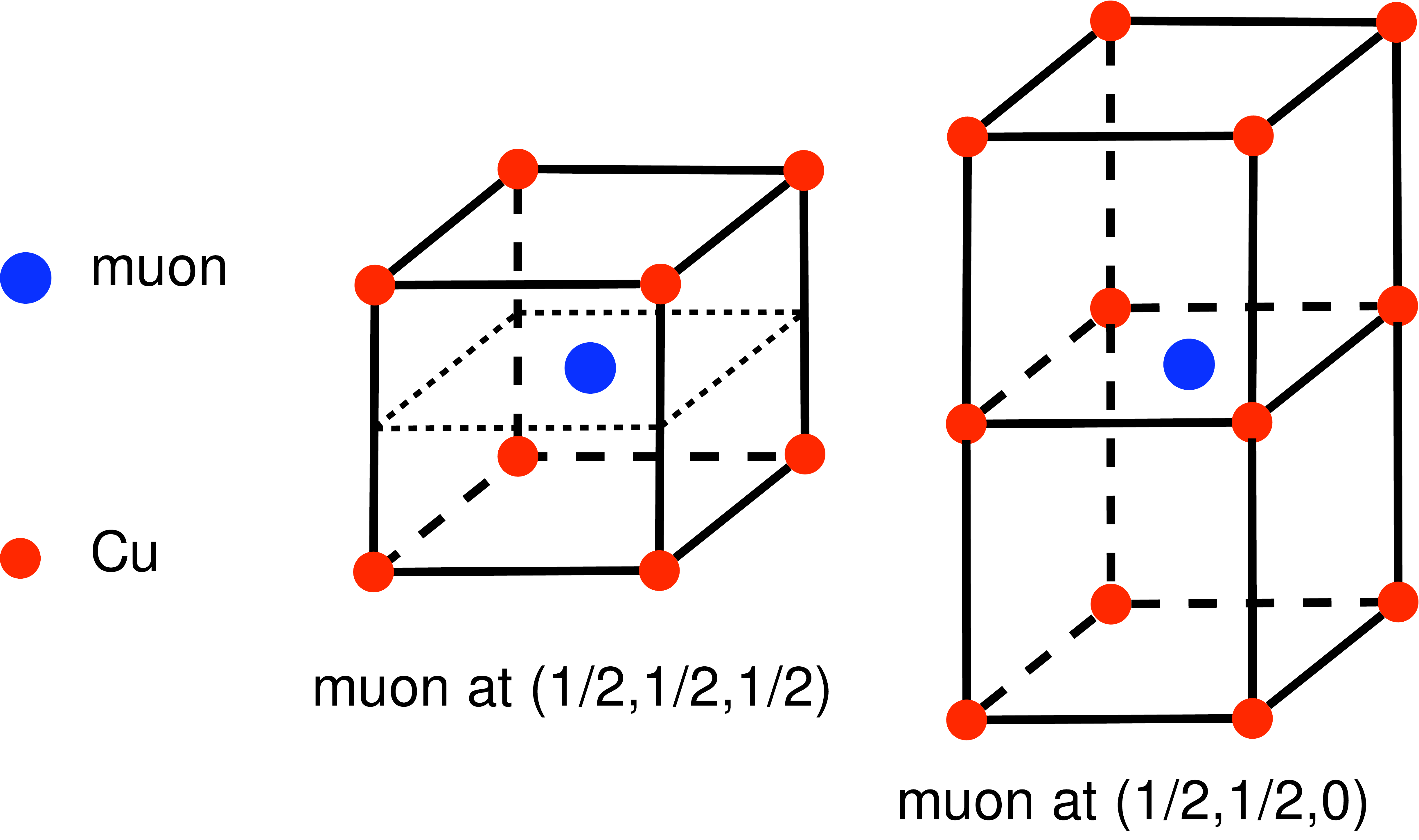}
\caption{\label{Fig:geometry}(Color online) Sketch of two cases considered in paper. Left panel: charged impurity (muon) placed at center of cube of transition-metal (Cu) sites. Right panel: impurity placed in plane. }
\end{figure} 

We suppose that the impurity is located at a position $R_\mu$ and has a charge $e$. We consider in detail two cases, shown in Fig.~\ref{Fig:geometry}. In one we take the impurity to be located in the center of a plaquette in  a $\mathrm{CuO}_2$ plane (i.e. the (1/2,1/2,0) position). In the other we place the impurity symmetrically between planes at  the $(1/2,1/2,1/2)$ position. Placing the impurity at these high symmetry points  allows us to use existing codes but as will be seen the physics we find is generic. 
 
We treat the screening using the self-consistent Hartree approximation.  The presence of the impurity potential changes the electronic  density on site $i$ (coordinate ${\vec R}_i$) from the average value ${\bar n}$ to a new value $n_i={\bar n}+\delta n_i$ so the total electrostatic potential is

\begin{equation}
V_i = -\frac{e^2}{\epsilon|\vec{R_i} - \vec{R_\mu}|}+
        \frac{e^2}{\epsilon}\sum_{j\neq i}\dfrac{\delta n_j}{|\vec{R_i} - \vec{R_j}|}
        \label{Vi}
\end{equation}
with $\epsilon$ the background dielectric constant.  The appropriate value of $\epsilon$ is not well established. Optical conductivity measurements \cite{Orenstein90} suggest that $\epsilon(\omega\rightarrow 0,q=0) \approx 4$, a value used in Ref.~\onlinecite{Shekhter08}, however what is needed is $\epsilon(\omega=0,q)$  for a range of  $q$  of the order of the shortest distance from a lattice site to the impurity position. Reasonable values of this quantity have not been determined. Calculations of the ``screened $U$'' for the related oxide material $\mathrm{SrVO}_3$ yield a high frequency unscreened $U\sim 14\mathrm{eV}$ and a low frequency screened quantity $W \approx 2\mathrm{eV}$ suggesting an electronic contribution to $\epsilon$ of $\epsilon \approx 7$. Lattice relaxation effects may increase the short scale  $\epsilon$ to a number of order $15$ (see, e.g. Ref.~\onlinecite{Okamoto06}]) but of course  lattice relaxations may induce other changes in the model. Resolving these uncertainties is beyond the scope of the paper; we have therefore performed our explicit calculations for  the two values  $\epsilon=4$ and $15$.

The remaining issue is the computation of $\delta n_i$ for a given distribution of $V_i$; this is discussed in the next section. Here we note that the scale of the screening effects is set by the dimensionless parameter
\begin{equation}
\gamma=\frac{e^2 }{\epsilon a}\frac{dn}{d\mu}
\label{gammadef}
\end{equation}
with $a$ the in-plane lattice parameter. We use the value  $a\approx 3.8\text{\AA}$ appropriate for cuprates.  The band theory estimate for the compressibility $dn/d\mu$ of weakly correlated electrons in the cuprate band structure is $dn/d\mu\approx 1.4/\mathrm{eV}$ so that $\gamma_\text{nonint}\approx  5/\epsilon$. As we shall see, for correlation strengths of the order of those believed to be relevant for high-temperature superconductors, the actual compressibility, and therefore the actual $\gamma$ are likely to be about an order of magnitude smaller. 

\section{Method}

\begin{figure}
\includegraphics[width=0.85\columnwidth]{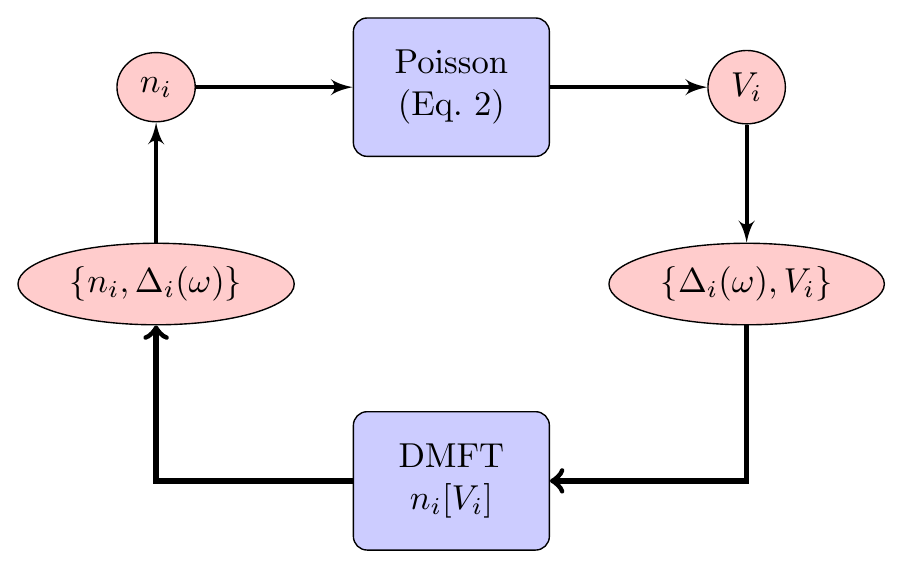}
\caption{\label{fig:DMFT_process}(Color online) Sketch of the self-consistency procedure used to calculate the charge density and hybridization functions in the vicinity of a charged impurity. Starting from an initial guess for the potential $V_i$ and the hybridization function $\Delta_i(\omega)$ the dynamical mean-field procedure is used to obtain a new density and hybridization function; the new density is used in Eq.~(\ref{Vi}) to obtain new potentials, and the process is iterated to self-consistency. Changes in the hybridization function are found to be sufficiently small that the DMFT loop may be solved once for the bulk material and $\partial n/\partial \mu$ obtained from this solution may then be used to update the density.}
\end{figure}

We require the solution of  a correlated electron problem in a spatially  inhomogeneous, self-consistently determined potential. There is no general and exact method for obtaining this information. We adopt here the single-site \cite{Georges96} and cellular \cite{Kotliar01} dynamical mean-field theory (CDMFT) approximations.  These methods capture important aspects of the strong correlation problem and, in particular, produce a Mott transition.  The single-site method is more computationally tractable; however the cluster method includes intersite  correlations and may provide a more reasonable picture of the spin dynamics. 

In the single-site DMFT method, the  electron self energy $\Sigma$ (in general a function of two coordinates and a frequency) is taken to be site local but may be different from site to site: $\Sigma(i,j;\omega)\rightarrow \Sigma_i(\omega)$. The self energy on each site $i$ is determined from the solution of a quantum impurity model (also different on each site). The impurity model is specified by the interaction $Un_\uparrow n_\downarrow$ and a hybridization function $\Delta_i(\omega)$. The impurity model Green's function on site $i$ is thus
\begin{equation}
G^\text{imp}_i(\omega)=\left[\omega+{\bar \mu}+V_i-\Delta_i(\omega)-\Sigma_i(\omega)\right]^{-1}.
\label{Gimp}
\end{equation}

The hybridization function is fixed by a self-consistency condition linking the Green's function $G^\text{imp}_i(\omega)$ of the impurity model on site $i$ to the $ii$ component of the lattice Green's function,
\begin{equation}
\begin{split}
G^\text{imp}_i(\omega)&=G^\text{lat}_i(\omega)\\
	&\equiv \left\{\left[\left(\omega+ 
{\bar \mu}+V_i-\Sigma_i(\omega)\right)\delta_{ij}-t_{ij} 
\right]^{-1}\right\}_{ii}.
\label{SCE}
\end{split}
\end{equation}

There are now two issues of self-consistency: $\Delta_i$ must be made self-consistent using Eqs.~(\ref{Gimp}) and (\ref{SCE}) and the potentials $V_i$ on all sites $i$ must be made self consistent with the computed densities (which are obtained from the $G^\text{imp}_i(\omega)$) using Eq.~(\ref{Vi}). To reach self-consistency one  begins with an initial guess for the site densities. From this one computes the $V_i$ via Eq.~(\ref{Vi}). Using these $V_i$ and an initial guess for the hybridization function one solves the DMFT equations, obtaining converged solutions for $G^\text{imp}_i$ and $\Delta_i$. From these we recompute $n_i$ and hence $V$ and continue the cycle until convergence is reached.  

Observe that the result of this procedure is that each  site has a hybridization function determined by neighboring sites, which have different densities. Thus a given site ``knows'' that it is in a spatially inhomogeneous environment, and therefore has properties which are different from those of a hypothetical bulk system in which all sites have a density equal to the density of the designated site. At various points in the ensuing discussion we will compare properties of a given lattice site $i$ with density $n_i$  to those that would be obtained in a bulk solid in which all sites had density $n_i$. 

\begin{figure}
\includegraphics[width=0.85\columnwidth]{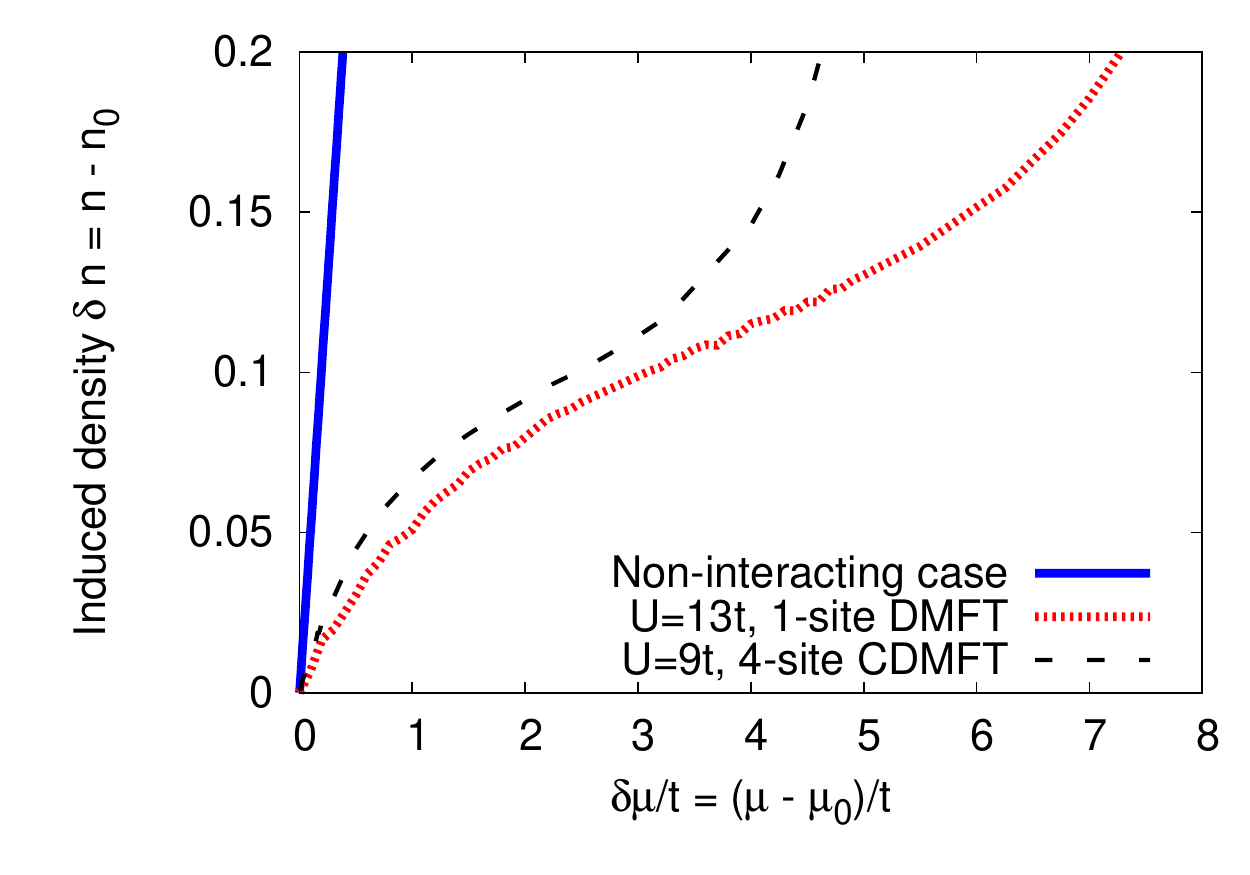}
\caption{\label{fig:dndmu}(Color online)  Local-density change  induced by  locally potential $\delta \mu_i$.  In all cases bulk density is $n=0.9$. Solid line: density change on site $i$, $\delta n_i$, induced by potential $\delta \mu_i$ applied on same site, computed for  non-interacting electrons using tight-binding  band parameters appropriate to high-temperature superconductors.  Dotted line:  same computation, but using single-site dynamical mean-field theory with $U=13t$ and hybridization parameters taken from calculation for $U=13t$ and $\delta=0.1$.  Dashed line: change in density on one site of a four-site plaquette induced by potential $\delta \mu$ applied to all four sites of plaquette, computed using CDMFT dynamical mean-field theory with $U=9t$ with hybridization function corresponding to $U=9t$ and $\delta=0.1$.}
\label{nofmu}
\end{figure} 

In practice the laborious procedure described above may be simplified. We have verified (for an example, see Fig.~\ref{fig:densityplots})  that to within an accuracy of $\sim 10\%$ the change in the hybridization function is negligible and the potential may be computed from the pure-system $n(\mu)$ curve. Thus we  use the homogeneous bulk hybridization function to compute the variation in local $n$ with local $\mu$  and use this to define the density/potential self-consistency with the result that it is necessary to solve the DMFT loop only once. The self-consistency procedure  is sketched in Fig.~[\ref{fig:DMFT_process}].

Figure ~\ref{fig:dndmu} shows the change in on-site density caused by a change in local chemical potential  computed for noninteracting electrons with bulk density corresponding to $0.1$ hole doping, and for two dynamical mean-field cases: single-site DMFT at $U=13t$ (slightly larger than the single-site DMFT critical $U$ which is $12t$ for this problem)  and cluster DMFT at $U=9t$ which is rather larger than the $6t$ needed to open a gap in this approximation but is in the range believed to be reasonable for cuprates.\cite{Comanac08} In the DMFT calculations the hybridization function was fixed at the form appropriate to a bulk material with density $n=0.9$. As expected, the correlation effects substantially reduce the local charge susceptibility: the initial slope is decreased by a factor of about $5$ relative to the noninteracting value and there is a substantial curvature. We also observe that even for $\delta n=0.1$, corresponding to a local density of one electron per site, the charge susceptibility remains non-vanishing (as expected because the local site is embedded in a metallic bath), whereas  the corresponding bulk system with density $n=1$ per site would be in a gapped phase with vanishing compressibility. 

The single-site method  is reasonably computationally tractable, enabling the exploration of a wider parameter space and a relatively detailed computation of physical quantities. However, this approximation overestimates the critical interaction required to drive the Mott transition, does not describe the ``pseudogap'' physics associated with underdoped cuprates and more generally does not capture the physics associated with short-ranged intersite correlations.  Cluster dynamical mean-field methods capture more aspects of cuprate physics, including a lower critical value for the Mott transition and some aspects of intersite spin correlations. They also exhibit a pseudogap.  However, the cluster methods are much more computationally expensive. Further, the widely used ``dynamical cluster approximation''\cite{Maier05}  is unsuited to impurity problems because it requires translation invariance. We therefore adopt the  CDMFT method \cite{Kotliar01} in which one tiles the lattice into real-space clusters; each cluster is regarded as a site of a new lattice of supercells. The hopping terms connecting sites in the same supercell are part of the cluster Hamiltonian while the hopping terms connecting sites on different clusters define the supercell band structure.  The new lattice is treated via single-site dynamical mean-field theory (albeit with a more complicated impurity), thus the self-consistency loop is the same as in the single-site case. We use four-site clusters. We solve the impurity model using the continuous-time quantum Monte Carlo method introduced in Ref.~\onlinecite{Werner06a}; for the four-site cluster we use the general (``matrix'') representation of Ref.~\onlinecite{Werner06b}. The method gives access both to the physical (lattice) electron Green's functions and to  correlation functions defined on the cluster model. While the cluster correlation functions are not identical to the corresponding lattice quantities, they are reasonable estimators of the physical correlators. 

One restriction should be noted: the impurity solver algorithm we use \cite{Werner06b} makes heavy use of symmetries and therefore requires that the four sites in the cluster have the same potential. Thus for cluster calculations we are limited to the case in which the impurity potential is the same for all four sites in the cluster. The geometry we use guarantees that this is the case for the four sites closest to the impurity on each plane. However, for the farther plaquettes, a problem arises, because one side of a plaquette is necessarily closer to the impurity than the other, so the local symmetry is broken. We treat this situation by solving the Poisson equation and then on each cluster replacing the potential by the average of the calculated potential over the cluster sites. The long range of the Coulomb interaction and the relatively small changes induced on farther neighbor clusters make this a reasonable approximation.

\section{Results: density distribution}

\begin{figure}
\includegraphics[width=0.85\columnwidth]{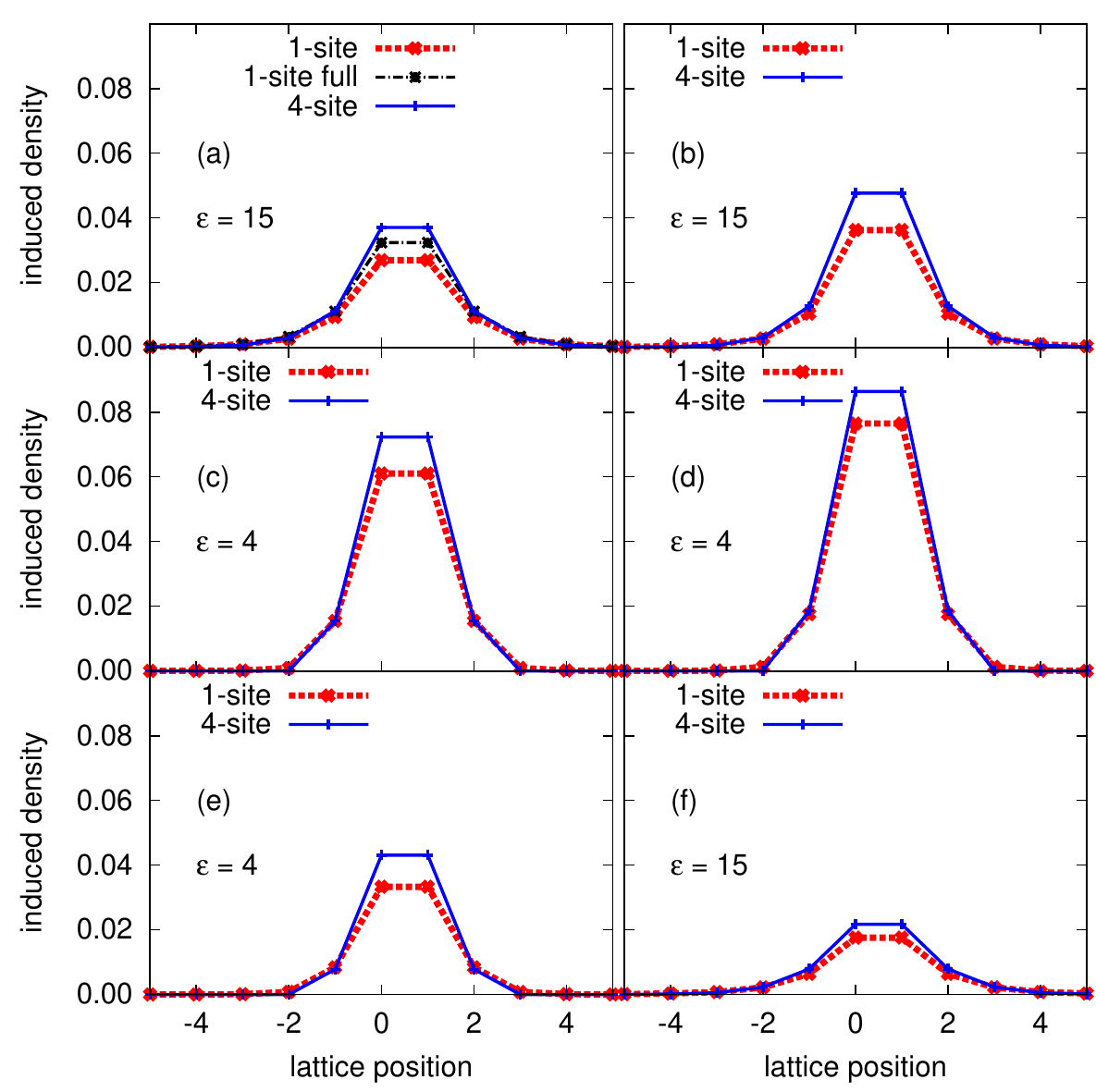}
\caption{\label{fig:densityplot}(Color online)  (a-d) Variation in conduction-band density per lattice site along line $(x,0,0)$  induced by impurities positioned at [(a) and (c)] $(1/2,1/2,1/2)$  and [(b) and (d)] $(1/2,1/2,0)$  calculated as described in the text using single-site DMFT with $U=13t$ and  and four-site DMFT with $U=9t$. [(e) and (f)] Variation in conduction-band density per lattice site along line $(x,0,1)$  induced by impurity positioned at $(1/2,1/2,0)$ for (e) $\epsilon=4$  and (f) $\epsilon=15$  calculated as described in the text using single-site DMFT with $U=13t$ and  four-site DMFT with $U=9t$.}
\label{fig:densityplots}
\end{figure}

An electrical conductor responds to a charged impurity by producing an electron density modulation (``screening cloud'') which screens the impurity charge. The  panels of Fig.~\ref{fig:densityplots} present the spatial distribution of the screening cloud induced by an impurity of charge $+1$ in a hole-doped superconductor. Shown is the charge density per lattice site along a line passing near to the impurity site    for two choices of background dielectric constant, $\epsilon=4$ and $\epsilon=15$, for two choices of impurity position (between planes or in the center of a plaquette in one plane) and for the two approximations we have used. We see that in all cases the density change is only appreciable on the sites adjacent to the impurity. For $\epsilon=4$ the density change on the sites nearest to the impurity is large enough to move the local density very close to the half filled value. For $\epsilon=15$ the density change is about a factor of two smaller than for $\epsilon=4$. The density profiles calculated for single-site and four-site DMFT  are very similar, because the density profile is controlled by the local compressibility which is similar for the two cases we have considered.  The density profiles calculated for the two impurity locations and for the farther plane are also similar because the $1/r$ variation in the unscreened Coulomb interaction is relatively slow. These calculations are performed using the simplified self-consistency loop described above; also shown are results obtained using the full DMFT procedure for the $\epsilon=15$, single-site DMFT case. The density changes induced by the impurity potential are seen to be generically of the order of $0.05$ electrons per site or less, which is less than but of the order of the doping for underdoped high-temperature superconductors.

\section{Results: spin correlations}

In this section we study how the screening cloud affects the local spin dynamics.   This is not  straightforward because the spin dynamics are expected to be strongly doping dependent in a homogeneous bulk system while  here we must treat a spatially inhomogeneous system.  We study impurity-model correlation functions, which can be directly measured in our simulations. These are not identical to the spin correlations of the actual lattice problem but are expected to have similar magnitude and similar doping and temperature dependence to those of the full lattice problem.  Further, the simulation gives results for Matsubara frequencies $\omega_n=2\pi nT$. The $n=0$ term is in essence the classical (thermal) part of the spin-spin correlation function while the $n\neq 0$ terms give some information on the quantum fluctuations in the system.  We present results for the spin correlations on the site nearest to the charge center and for the second neighbor, and compare the results to those found far from the charge center and also to those computed for a hypothetical bulk system  with average density equal to that on the site  nearest to the charge center.

\begin{figure}[ht]
\includegraphics[width=0.85\columnwidth]{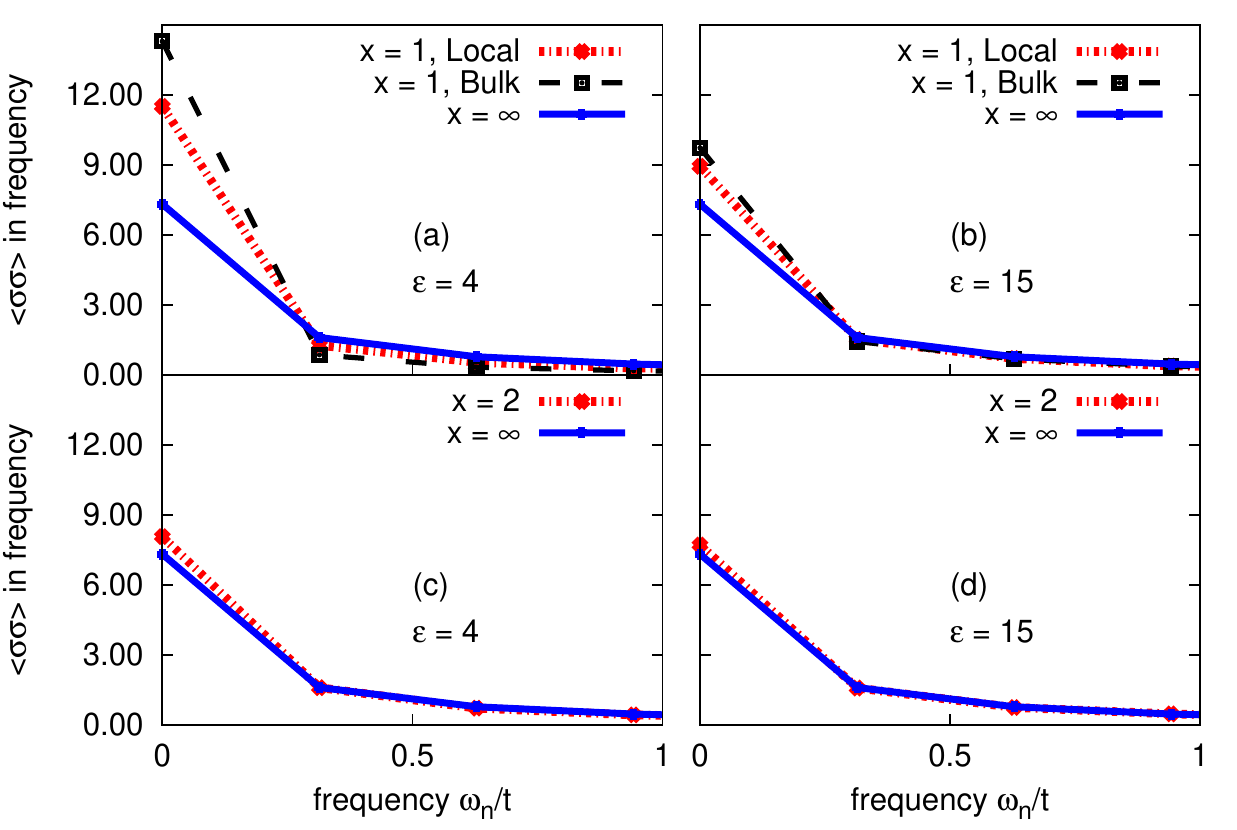}
\caption{(Color online) Impurity-model spin-spin correlation along line $(x,0,0)$ computed from single-site dynamical mean-field theory as function of Matsubara frequency at inverse temperature  $\beta = 20$ and dielectric constants indicated for charged impurity at position $(1/2,1/2,0)$. Upper panels are onsite correlators for sites nearest to muon $(x = 1)$, lower panels are onsite correlators for second neighbors to muon $(x = 2)$, correlator for site far from muon is also included.}

\label{fig:spincorr1site}
\end{figure} 

Figure~\ref{fig:spincorr1site} shows the impurity-model spin-correlation functions computed using single-site dynamical mean-field theory.  We see that the ``classical'' (zero Matsubara frequency) spin correlations of sites near the muon are enhanced relative to the value far from the muon site but are not as large as those of a hypothetical system with average density equal to the density on the near-muon sites. This shows that the spin correlations on a given site are controlled not only by the density on the site but also by the properties of the neighboring sites. For the case $\epsilon=4$ we see that the changes are substantial (increasing the value at the lowest Matsubara frequency by a factor of about $1.5$, which in turn is about half of the increase that would occur in a sample whose average density was set equal to the density on the impurity site). On the other hand for $\epsilon=15$ the changes, although visible, are much smaller.

\begin{figure}[ht]
\includegraphics[width=0.85\columnwidth]{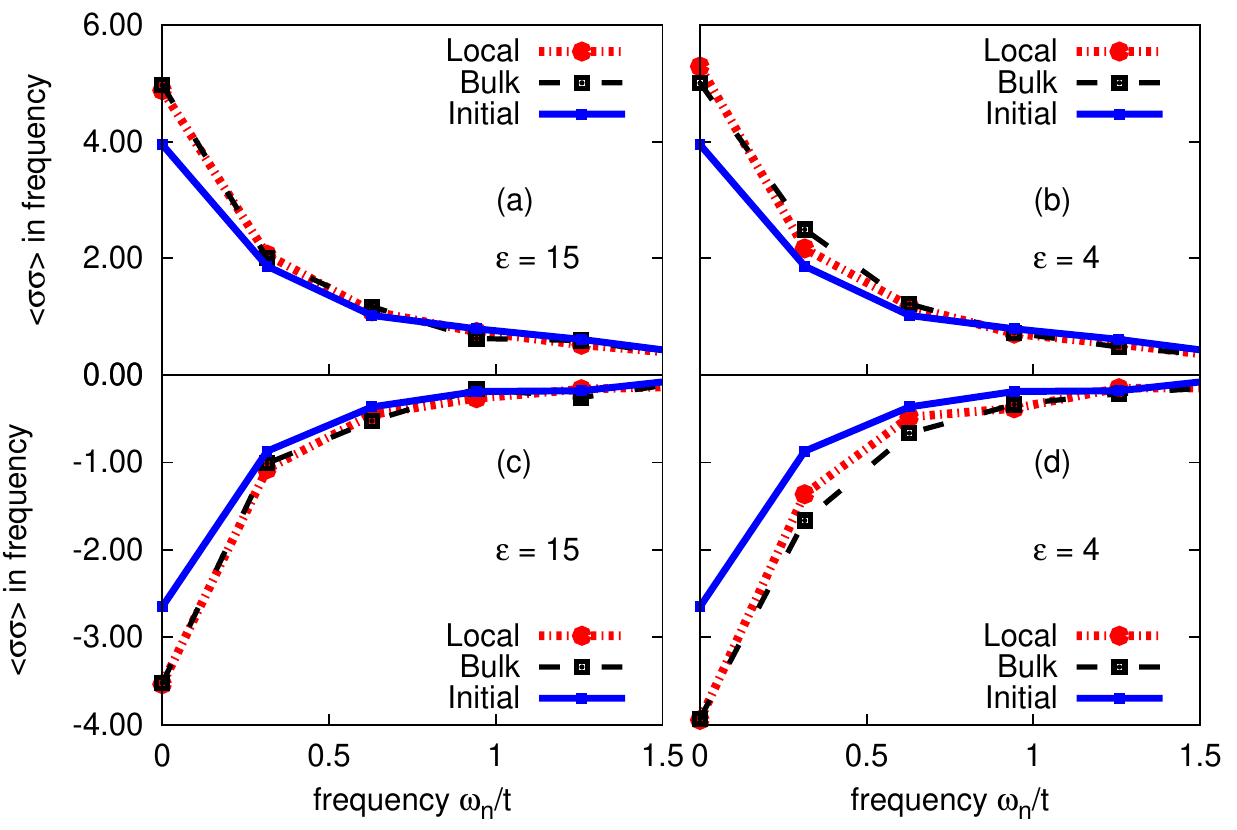}\\
\caption{(Color online) On-site (upper panels) and first-neighbor (lower panels) impurity model spin-spin correlation  computed from four-site CDMFT cluster dynamical mean-field theory as function of Matsubara frequency at inverse temperature  $\beta = 20$ and dielectric constants indicated, for charged impurity at position $(1/2,1/2,0)$. Solid line with solid squares (blue online) shows  correlator computed for sites far from impurity site.  Dotted line with solid circles (red online) shows correlator computed on designated site. Dashed line with open squares (blue online) shows correlated computed for bulk system with mean density equal to density on impurity site.}

\label{fig:spincorr4site}
\end{figure}

The single-site dynamical mean-field theory is known to provide a poor approximation to the spin correlations of a doped Mott insulator, at least in two spatial dimensions, because it neglects antiferromagnetic correlations.  We have therefore also considered the spin correlations in the four-site CDMFT calculations. Here the difference between on-site and first-neighbor impurity-model spin correlations reveals the importance of antiferromagnetic fluctuations.   Figure \ref{fig:spincorr4site} shows results for the same parameters as in Figure \ref{fig:spincorr1site}. We see that the first-neighbor correlations increase by about $40\%$ for $\epsilon=4$ and about $25\%$ for $\epsilon=15$. We have also examined other correlation functions, in particular the equal-time singlet-pair correlations which are the dominant fluctuations on the four-site plaquette, finding that these are enhanced by similar amounts.

\section{Conclusion}

In this work, we have shown how ``strong correlation''  effects alter the response of a material to a local charge inhomogeneity. We introduced a general method, based on dynamical mean-field theory, for calculating these effects and applied it to the question of the changes produced by the presence of a muon in a high-temperature superconductor.  We found, in qualitative agreement with previous work, \cite{Shekhter08} that the muon is not a ``soft probe:'' although the main correlation effect is a suppression of the charge susceptibility by a factor of 3-4 relative to band theory,  the charge field associated with the muon may  produce a significant change in the charge density on nearby sites, of order $~0.05$ electrons per site. In a less strongly correlated material, the change in charge density on the near-muon sites would be larger.  A crucial issue in determining the scale of the effects was found to be the value of the static, short wavelength, dielectric constant. The effect on the local spin correlations is  smaller than the effect on the charge density but is not negligible.

Our calculation involves several approximations. The most important is the value of the dielectric constant, which is not known {\it a priori}. Varying the dielectric constant over the range of values which have been proposed for   cuprates or other oxides leads to factor of 2 changes in our results. A closely related issue concerns the local lattice distortions which would normally be induced near a charged impurity (see Ref.~\onlinecite{Okamoto06} for an example in a different context). These have the potential to change local hopping amplitudes and perhaps correlation strengths, although correlation strengths, being an atomic property, will be less strongly affected. In view of the importance of muon-spin rotation as a probe of correlated materials, these issues deserve further investigation, perhaps via a band-theory calculations.  A second approximation is the use of the dynamical mean-field method.  The consistency of our single-site and four-site results for the charge compressibility suggest that our basic findings for the density correlations are a reasonable estimate of the correct behavior. However, the calculated spin correlations are probably subject to larger uncertainties, which are at this point not easily quantified. We know that the charge perturbation is important only on sites immediately adjacent to the charge center. The spin correlations on these sites, which are the ones which would be probed by a muon, are affected  both by the on-site property (the change in local density) and by the properties of the nearby sites, and the nearby sites in turn both affect and are affected by the near-impurity sites. If the intersite spin coupling is strong it is possible that the magnetic properties are controlled by the sites farther away from the muon.  It is very likely that the dynamical mean-field methods we use underestimate these spin-correlation effects.  Their investigation is an important open problem.

While our specific numerical results  were obtained for  model parameters appropriate to high-temperature superconductors, they have  implications for the more general issue of the  response  of a correlated electron material to a charged impurity.  To illustrate this point we consider a generic correlated material, which we assume to be a more or less cubic lattice. For simplicity place the charged impurity at the center of a cube of sites. The charged impurity will induce a screening cloud containing one electron. The length scale over which this charge is distributed is set by the density-density correlation function of the charged material and the background dielectric function. If values typical of a weakly correlated material are used and the dielectric function is of the order of $10$ or less, a simple extension of the estimates we have presented indicates that almost  the entire screening charge sits as close to the impurity as it can get. The density change on the near-impurity sites would then (for the simple cubic situation we have considered) be approximately $1/6$ electron per site, a change large enough to affect the local physics.   On the other hand, if strong correlation effects are important (as in the case of high-$T_c$ materials where they reduce the charge response by a factor of $5$ or more), the total charge would be spread over a wider range and the concentration on the near-impurity sites would be substantially smaller.  However, the relative effect on the local physics would still not be small, as the  greatest suppression of charge response occurs for a lightly doped Mott insulator, where the important scale is the doping, which would itself be small. Thus even in this case we would expect that a charged impurity would change the local physics noticeably.  
A quantitative test of our theory would involve measurements of the near-impurity charge-density profile and a comparative measurement of spin dynamics near to and far from the impurity site.

Our results have implications for muon-spin-resonance experiments on transition-metal oxides. Muons are an important probe of the spin dynamical of condensed matter physics, but a muon has a charge $+1$, and the results presented here indicate that in transition metal oxides a muon is unlikely to be a ``soft'' probe; rather, it significantly perturbs the medium in which it is embedded.
Feyerherm {\it et al.}\cite{Feyerherm95} reached a similar conclusion in a study of $\mathrm{PrNi}_5$, a  rare earth system with more complicated physics, showing that muons significantly perturb the crystal-field structure on the near muon $\mathrm{Pr}$ sites. In the case studied here we showed that the perturbation due to the muon affects local properties such as the near-muon spin dynamics and presumably (although we have not investigated this) the size of the ordered moment. It is important to note that a dilute concentration of muons should have only negligible effects on ``global'' or long-range properties such as magnetic phase boundaries or superconducting penetration depth.

{\it Acknowledgements:} We thank  W.~Hardy for bringing the issue to our attention, W.~Hardy, G.~Luke, J.~Sonier, and Y.~Uemura for discussions and the Canadian Institute for Advanced Research for enabling the interaction.  We acknowledge support from NSF-DMR-0705847. A.J.M. and E.G. also acknowledge the hospitality of the Kavli Institute for Theoretical Physics, enabled by  partial  support from the  National Science Foundation under Grant No. PHY05-51164.  H.T.D. acknowledges partial support from Vietnam Education Foundation (VEF). The calculations were performed using a code partly written by P. Werner and  based on the  ALPS library (Ref.~\onlinecite{Albuquerque20071187}).

\end{document}